\documentclass[11pt]{article}

\usepackage[utf8]{inputenc}
\usepackage[T1]{fontenc}
\usepackage{lmodern}
\usepackage[margin=1in]{geometry}
\usepackage{amsmath,amssymb}
\usepackage{graphicx}
\usepackage{booktabs}
\usepackage{longtable}
\usepackage{array}
\usepackage{multirow}
\usepackage{tabularx}
\usepackage{makecell}
\usepackage{url}
\usepackage{textcomp}
\usepackage{hyperref}
\usepackage[numbers]{natbib}
\usepackage{calc}

\makeatletter
\@ifundefined{c@none}{}{}
\makeatother

\providecommand{\tightlist}{%
  \setlength{\itemsep}{0pt}\setlength{\parskip}{0pt}%
}

\title{BattMo - Battery Modelling Toolbox}
\author{\parbox{0.98\textwidth}{\centering Xavier Raynaud\textsuperscript{*,1}, Halvor Møll Nilsen\textsuperscript{1}, August Johansson\textsuperscript{1}, Eibar Flores\textsuperscript{2}, Lorena Hendrix\textsuperscript{2}, Francesca Watson\textsuperscript{1}, Sridevi Krishnamurthi\textsuperscript{2}, Olav Møyner\textsuperscript{1}, Simon Clark\textsuperscript{2}\\[0.75em]
{\small \textsuperscript{1}SINTEF Digital, Dept. of Mathematics and Cybernetics, Norway}\\{\small \textsuperscript{2}SINTEF Industry, Dept. of Sustainable Energy Technology, Norway}\\{\small \textsuperscript{*}Corresponding author}}}
\date{}

\begin{document}

\maketitle

\begin{abstract}
This paper presents the Battery Modelling Toolbox (BattMo), a flexible finite volume continuum modelling framework in MATLAB\textsuperscript{\textregistered} \citep{MATLAB} for
simulating the performance of electro-chemical cells. BattMo can quickly set up and solve models for a variety of battery
chemistries, even considering 3D designs such as cylindrical and prismatic cells.

The simulation input parameters, including the material parameters and geometric descriptions, are specified through
JSON schemas. In this respect, we follow the guidelines of the Battery Interface Ontology (BattINFO) to support semantic
interoperability in accordance with the FAIR principles \citep{fair}.

The Doyle-Fuller-Newman (DFN) \citep{Doyle1993ModelingCell} approach is used as a base model. We include fully coupled
thermal simulations. It is possible to include degradation mechanisms such as SEI layer growth, and the use of composite
material, such as a mixture of Silicon and graphite.

The models are set up in a hierarchical way, for clarity and modularity. Each model corresponds to a computational graph,
which introduces a set of variables (the nodes) and functional relationship (the edges). This design enables the
flexibility for changing and designing new models.

The solver in BattMo uses automatic differentiation and support adjoint computation. We can therefore compute the
derivative of objective functions with respect to all parameters efficiently. Gradient-based optimization
routines can be used to calibrate parameters from experimental data.

\end{abstract}

\section{Statement of need}\label{statement-of-need}

New high-performance electro-chemical systems such as Li-ion and post-Li-ion batteries are essential to achieve the goals of the electric energy transition. Developing
rigorous digital workflows can help industrial and research institutions reduce the need for physical prototyping
and derive greater insight and knowledge from their data.

BattMo extends this effort by supporting fully 3D geometry and the
possibility to easily modify the underlying equations. We provide a
library of standard parameterized battery geometries. Design
optimization can also be done on the geometry, which is an essential
part of the design. BattMo is useful for both experienced battery
designers, wanting to optimize and virtually test different designs,
research software developers integrating simulation tools in
electro-chemical system workflows as well as beginners wanting to
understand fundamental processes.

A challenge for physics-based models for batteries and other electro-chemical systems is the difficulty to calibrate the parameters. With an adjoint-based
approach, we can effectively calibrate the models from experiments in a reasonable computational time.

\section{State of the field}\label{state-of-the-field}

Recently, a variety of
open-source battery modelling codes have been released including PyBaMM \citep{sulzer2021python}, cideMOD
\citep{CiriaAylagas2022}, LIONSIMBA \citep{torchio2016lionsimba}, and PETLion \citep{Berliner_2021}, among others. These
open-source frameworks help the community reduce the cost of model development and help ensure the
validity and the reproducibility of findings. PyBaMM is, to our knowledge, the most popular option. Unlike BattMo, it has just recently (2025) allowed for simulations in 2D and 3D, and for calibration and optimization, it is usually recommended to utilize PyBOP \citep{Planden2025} which is an external package (although coupled tightly to PyBaMM). BattMo's foundation is designed for coupled electro-chemical-thermal simulations in 3D, as well calibration and optimization. The extensive portfolio of degradation models in PyBaMM is a great strength. In BattMo we are currently allowing for SEI (Safari \citep{Safari2009} and Bolay \citep{Bolay2022} models) as well as plating \citep{Hein2020}. Based on FEniCS \citep{fenics}, cideMOD by design allows for simulations in 3D, but the lack of automatic differentiation means handling complex nonlinear systems and performing design optimization is more challenging.

\section{Software design}\label{software-design}

BattMo builds on the MATLAB Reservoir Simulation Toolbox \citep{mrst-book-i} which provides the foundation for meshing
intricate geometries, solving large nonlinear systems of equations, visualizing the results, etc, in addition to fundamental robust, low-order finite volume methods and fully implicit time discretization. Although developed in
MATLAB\textsuperscript{\textregistered}, we try to provide Octave \citep{octave} compatibility. Neither BattMo nor MRST rely on extra MATLAB\textsuperscript{\textregistered} packages; the basic license is sufficient. We do recommend using AMG preconditioners from the open-source
AMGCL library \citep{Demidov2020} for fast, multi-threaded solution of linear systems.

\subsection{Functionality overview}\label{functionality-overview}

The main features of BattMo are summarized in the following list:

\begin{itemize}
\tightlist
\item
  JSON based input with schema
\item
  Library of parameterized battery formats
\item
  Flexible graph-based model design
\item
  Fully coupled electro-chemical--thermal models
\item
  3D visualization
\item
  Parameter calibration
\item
  Design optimization
\item
  Support for standard protocols such as CC, CV, CCCV, and time series
\item
  SEI layer growth model
\item
  Composite material model
\item
  Silicon swelling model
\item
  Material database for NMC, NCA, LFP, LNMO, SiGr, electrolytes, etc
\item
  Alkaline membrane electrolyser model
\item
  Proton ceramic membrane model
\end{itemize}

\subsection{Battery format library}\label{battery-format-library}

We support coin cells, jelly roll cells and multi-pouch cells with different tab layouts. The geometries are parameterized and can be modified
using a simple set of parameters. 1D and 2D grids for P2D and P3D models can also be generated.

\begin{figure}
\centering
\includegraphics[width=1\linewidth,height=\textheight,keepaspectratio,alt={A selection of the parameterized battery geometries available. Clockwise from top left are: A single-layer pouch cell, CR 2016 coin cell, 30-layer pouch cell and jelly roll cylindrical cell. }]{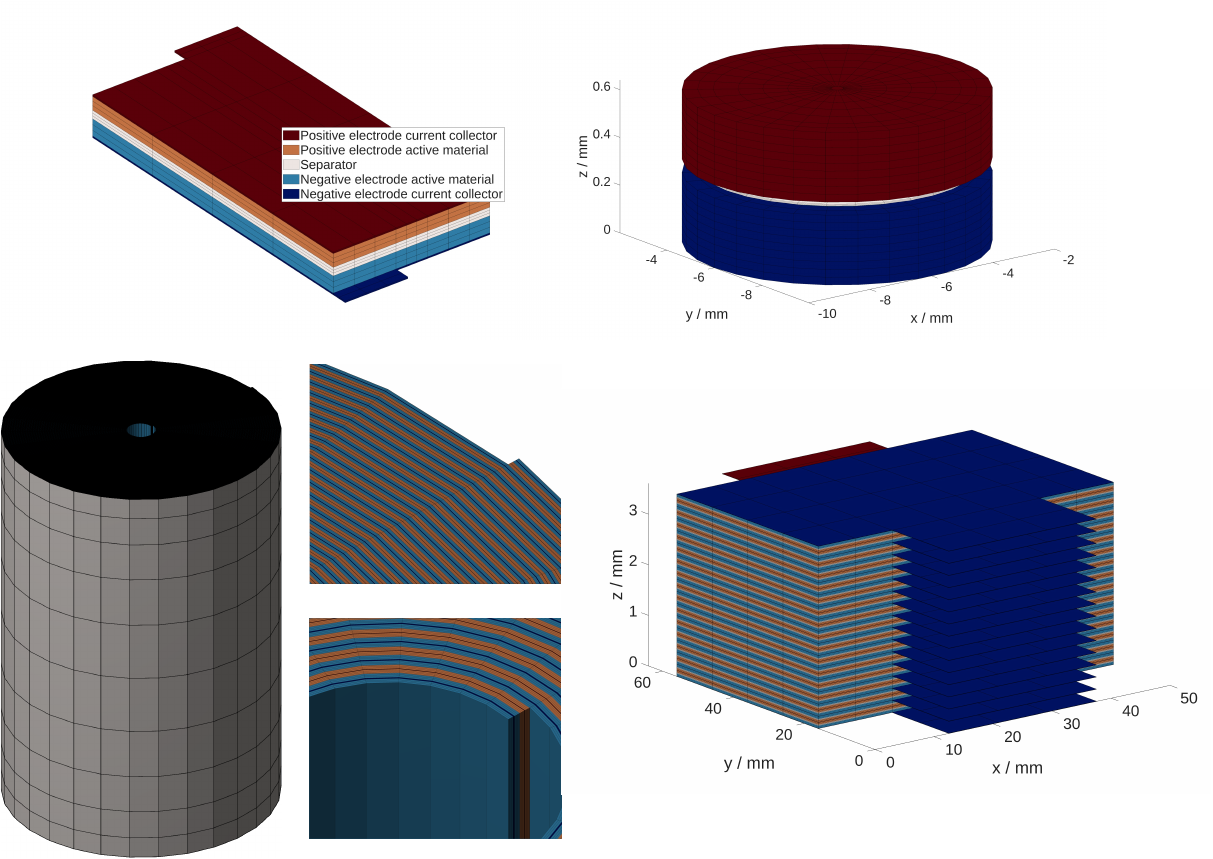}
\caption{A selection of the parameterized battery geometries available. Clockwise from top left are: A single-layer pouch cell, CR 2016 coin cell, 30-layer pouch cell and jelly roll cylindrical cell. \label{fig:geometries}}
\end{figure}

\subsection{Graph based model development}\label{graph-based-model-development}

BattMo has the ambition to support a variety of electro-chemical systems. The complexity of such models increases rapidly as models are extended or coupled. To manage this, BattMo introduces a computational graph-based model design. Each model is defined as a graph whose nodes represent variables and whose directed edges represent their functional relationships. A model is ready for simulation when the graph's roots are the governing variables and its leaves the governing equations. Interactive tools are available to explore these graphs.

Model hierarchy is an essential part of the framework. Coupling two models is done by creating a new coupling model that contains both as sub-models. Their graphs become sub-graphs, and new edges are added to represent coupling mechanisms, allowing most sub-models to remain unchanged. For more details, see the \href{https://battmo.org/BattMo/computationalGraph/graphdoc.html}{documentation}.

\begin{figure}
\centering
\includegraphics[width=0.8\linewidth,height=\textheight,keepaspectratio,alt={The computational graph of an active material model. }]{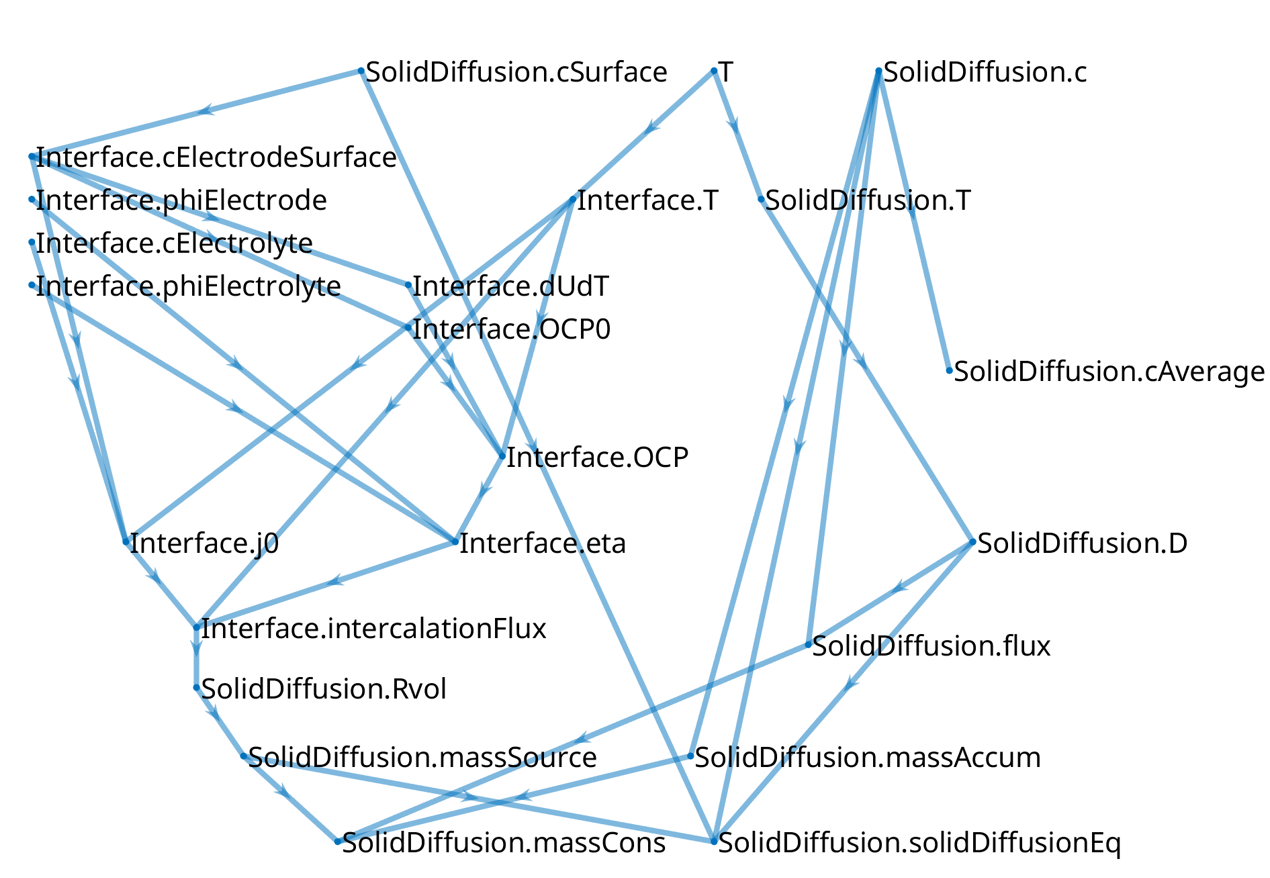}
\caption{The computational graph of an active material model. \label{fig:graph}}
\end{figure}

\subsection{Examples}\label{examples}

Numerous documented and tested examples are provided with the code and demonstrate the features listed above. We seek to be cross-compatible, allowing for any chemistry and material combination with any battery format. A selection of the examples are documented at \url{https://battmo.org/BattMo/}. The complete list of examples are available at \url{https://github.com/BattMoTeam/BattMo/tree/main/Examples}.

\subsection{BattMo family}\label{battmo-family}

The following software include the BattMo family:

{\def\LTcaptype{none} 
\begin{longtable}[]{@{}
  >{\raggedright\arraybackslash}p{(\linewidth - 2\tabcolsep) * \real{0.4390}}
  >{\raggedright\arraybackslash}p{(\linewidth - 2\tabcolsep) * \real{0.5610}}@{}}
\toprule\noalign{}
\begin{minipage}[b]{\linewidth}\raggedright
Software
\end{minipage} & \begin{minipage}[b]{\linewidth}\raggedright
Description
\end{minipage} \\
\midrule\noalign{}
\endhead
\bottomrule\noalign{}
\endlastfoot
\href{https://github.com/BattMoTeam/BattMo}{BattMo} & MATLAB\textsuperscript{\textregistered} version presented in this publication \\
\href{https://github.com/BattMoTeam/BattMo.jl}{BattMo.jl} & Julia version \\
\href{https://github.com/BattMoTeam/PyBattMo}{PyBattMo} & Python wrapper around BattMo.jl \\
\href{https://app.battmo.org/}{BattMoApp} & Web-application built on top of BattMo.jl \\
\end{longtable}
}

\section{Research impact statement}\label{research-impact-statement}

By combining a modular modeling framework with documented JSON-based
inputs, tutorials, and interfaces such as the Julia bridge, BattMo
provides reproducible battery cell development for research and
education. BattMo has been used in several commercial and
non-commercial projects. Publications include \citep{Clark2026} and
\citep{Schmitt2026}. It has been important for the success of several EU
projects, primarily HYDRA (875527) and BATMAX (101104013), where it
has been used to model novel LNMO cells, as well as commercial
LFP and NMC cells in both 1D and 3D.

\section{AI usage disclosure}\label{ai-usage-disclosure}

Development of BattMo started before AI tools were
introduced. Therefore only very minor use of AI tools (ChatGPT from
OpenAI and Microsoft Copilot) were used in the development of this
software. AI tools have been used to assist phrasing and conciseness
when writing this manuscript. In all instances, the authors have
reviewed the suggestions and edited them if necessary.

\section{Acknowledgements}\label{acknowledgements}

BattMo has been mainly been developed and used in projects funded by the European Union. We acknowledge contributions from the European Union, Grant agreements 101069765 (IntelLiGent), 875527 (HYDRA), 957189 (BIG-MAP), 101104013 (BATMAX) and 101103997 (DigiBatt).

\nocite{*}

\bibliographystyle{plainnat}
\bibliography{paper}

\end{document}